\documentclass[prb,twocolumn,showpacs,preprintnumbers,amsmath,amssymb]{revtex4}

\usepackage{graphicx}
\usepackage{dcolumn}
\usepackage{bm}

\begin{document}

\title{Subharmonic gap structure in short ballistic graphene junctions}

\author{J.C. Cuevas$^{1,2,3}$ and A. Levy Yeyati$^{1}$}
\affiliation{$^1$Departamento de F\'{\i}sica Te\'orica de la Materia Condensada,
Universidad Aut\'onoma de Madrid, E-28049 Madrid, Spain \\
$^2$Institut f\"ur Theoretische Festk\"orperphysik, Universit\"at Karlsruhe,
D-76128 Karlsruhe, Germany \\
$^3$Forschungszentrum Karlsruhe, Institut f\"ur Nanotechnologie, D-76021
Karlsruhe, Germany}

\date{\today}

\begin{abstract}
We present a theoretical analysis of the current-voltage characteristics of a 
ballistic superconductor-normal-superconductor (SNS) junction, in which a strip 
of graphene is coupled to two superconducting electrodes. We focus in the
short-junction regime, where the length of the strip is much smaller than superconducting
coherence length. We show that the differential conductance exhibits a very rich
subharmonic gap structure which can be modulated by means of a gate voltage. On approaching
the Dirac point the conductance normalized by the normal-state conductance is identical to
that of a short diffusive SNS junction.
\end{abstract}


\maketitle

\emph{Introduction.--}
Recently it has become possible to manipulate graphene, a single atomic layer of carbon, 
and to establish electrical contacts with it~\cite{Novoselov2004,Novoselov2005,Zhang2005}. 
This has opened the possibility of studying its transport properties, which are determined 
by the dynamics of two-dimensional Dirac fermions (massless excitations governed by a 
relativistic wave equation)~\cite{Novoselov2005,Zhang2005}. Moreover, the fabrication of a 
new class of hybrid structures in which superconductors are coupled via graphene is now 
feasible~\cite{Heersche2006}. It has already been shown theoretically that the 
nature of the low-lying graphene excitations might lead to unexpected features in transport
properties of these heterostructures~\cite{Beenakker2006,Titov2006}. Thus for 
instance, Titov and Beenakker~\cite{Titov2006} have recently demonstrated that
a SNS junction with an undoped strip of graphene as a normal region can sustain a
supercurrent and it exhibits an unusual ``quasi-diffusive" scaling of the dc
Josephson effect.

The supercurrent in graphene hybrid systems may nicely reveal the interplay 
of superconductivity and the relativistic quantum dynamics of the electrons in
these carbon structures. However, a direct comparison of the supercurrent between
theory and experiment is usually not straightforward. This is due to the fact
that the superconducting phase is prone to both quantum and thermal fluctuations,
which depend on the electromagnetic environment in which the junction is 
embedded~\cite{Joyez1999}. Thus, unless the environment is carefully designed, which
is not an easy task, the supercurrent might be greatly reduced as compared with the 
theoretical predictions for idealized situations. For this reason, we propose to look at the 
current-voltage characteristics (I-V) of graphene SNS contacts, which do not suffer 
from the problems just mentioned. The main feature of the I-V curves of SNS 
junctions is the appearance of the so-called \emph{subharmonic gap structure}, 
which consists of a series of conductance maxima at voltages $2\Delta/ne$, where 
$n$ is an integer and $\Delta$ is the energy gap of the electrodes. This structure 
originates from the ocurrence of multiple Andreev reflections (MARs)~\cite{Klapwijk1982}. 
The microscopic theory of these tunneling processes developed in the 
1990's~\cite{Bratus1995,Cuevas1996} has been shown to describe accurately
the I-V characteristics of superconducting atomic point-contacts~\cite{Scheer1997}.

In this Communication we present a theoretical study of the I-V characteristics of SNS junctions,
where the normal region is a ballistic strip of graphene. We consider the experimentally
relevant case of a short junction, in which the length of the strip is smaller than
the superconducting coherence length $\xi$. We show that the conductance exhibits a
pronounced subharmonic gap structure that can be tuned with a gate voltage. In the
case of zero gate voltage, i.e. at the Dirac point, the I-V curves are, after rescaling,
identical to those of a short diffusive SNS junction, which is a new consequence of the 
relativistic dynamics of the electrons in graphene. 

\begin{figure}[t]
\begin{center}
\includegraphics[width=0.8\columnwidth,clip]{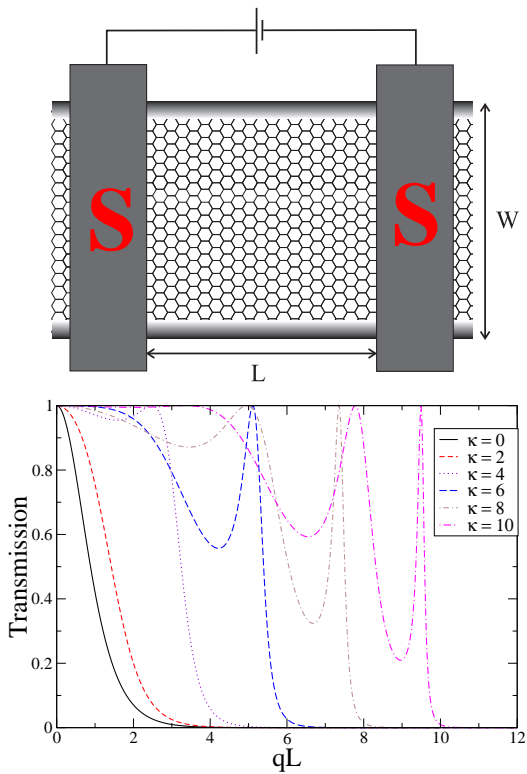}
\caption{\label{setup} (Color online) Upper panel: Schematic of a strip of graphene 
of width $W$, contacted by two superconducting electrodes (black rectangles) 
at a distance $L$. A voltage source drives a current through the strip. 
A separate gate electrode (not shown) allows the carrier concentration in 
the strip to be tuned around the neutrality point. Lower panel: Normal transmission 
of the conduction channels of the graphene junction as a function of the transversal 
momentum in the limit $W/L \gg 1$. The different curves correspond to different values 
of the dimensionless gate voltage $\kappa = e |V_{\rm gate}|L/ \hbar v$.} 
\end{center}
\end{figure}

\emph{Model for the graphene junction.--}
We consider the system shown schematically in Fig.~\ref{setup} (upper panel), where
a graphene strip is coupled to two superconducting electrodes. We also
assume that an additional gate electrode allows to control the carrier concentration
in the normal part of the junction. We focus here in the case where the width of the 
graphene strip $W$ is much larger than the junction length $L$ (in this case the details 
of the microscopic description of the strip edges become irrelevant). Our main goal
is the analysis of the superconducting I-V characteristics in the experimentally most 
relevant \emph{short-junction} regime where $L$ is small relative to $\xi$. 
In terms of energy scales, this condition requires $\Delta \ll \hbar v/L$, where $v$ is the
energy-independent electron velocity in graphene. It has been shown that, as long as
the normal transmission coefficients do not depend on energy on the scale of $\Delta$,
the transport properties of a short SNS junction can be expressed as a sum of independent 
contributions from individual conduction modes~\cite{Bardas1997}. Therefore, the transport
properties depend only on the distribution of transmission probabilities $\tau_i$ of
these modes in the normal state. For describing the transmission distribution of a
graphene junction one can adopt the model introduced in Ref.~\onlinecite{Tworzydlo2006},
which is briefly discussed in the next paragraphs.

The authors of Ref.~\onlinecite{Tworzydlo2006} considered the junction depicted in 
Fig.~\ref{setup} (upper panel). They determined the electron wave functions by solving
the Dirac equation for massless fermions. In particular, they showed that assuming 
``infinite mass" boundary conditions at the strip edges $y=0$ and $y=W$, the 
transversal momenta are quantized as 
\begin{equation}
\label{qc}
q_n=\frac{1}{W}\pi\left(n+\tfrac{1}{2}\right),\;\; n=0,1,2, \dots ,
\end{equation}
with $n$ labeling the modes. Each mode has a twofold valley degeneracy.

In the model of Ref.~\onlinecite{Tworzydlo2006}, the gate voltage enters in the Dirac 
equation as an electrostatic potential. This potential $V(x)=V_{\rm gate}$ for $0<x<L$  
determines the concentration of the carriers in the strip. The value $V_{\rm gate}=0$ 
corresponds to charge neutrality, being the point where electron and hole excitations 
are degenerate (known as the Dirac point). The electrodes are modelled by taking a large 
value $V(x)=V_{\infty}$ in the leads $x<0$ and $x>L$. (The parameter $V_{\infty}$ drops
out of the results, if $|V_{\infty}|\gg |V_{\rm gate}|$.) The number of propagating
modes $N$ is given by $N={\rm Int}\,\bigl(k_{\infty}W/\pi+\tfrac{1}{2}\bigr)$, 
with $e|V_{\infty}|=\hbar vk_{\infty}$. We are interested in the limit $|V_{\infty}|
\rightarrow\infty$ of an infinite number of propagating modes in the leads. By 
matching the solutions of the Dirac equation in the three regions of the junction one
finds that the transmission probabilities at the Fermi level are given by~\cite{Tworzydlo2006}
\begin{equation}
\tau_{n} = \left|\frac{2\delta^{2}-2(q_{n}-k_{n})^{2}}{e^{k_{n} L}(q_{n}-k_{n}+i\delta)^{2}
+e^{-k_{n} L}(q_{n}-k_{n}-i\delta)^{2}}\right|^{2},\label{Tna}
\end{equation}
with $\delta=e|V_{\rm gate}|/\hbar v$ and $k_{n}=\sqrt{q_{n}^{2}-\delta^{2}}$ for 
$q_{n}>\delta$ or $k_{n}=i\sqrt{\delta^{2}-q_{n}^{2}}$ for $q_{n}<\delta$.
These transmssion coefficients are plotted in Fig.~\ref{setup} (lower panel) for
different values of the dimensionless gate voltage $\kappa = e |V_{\rm gate}|L/ \hbar v$.

The normal-state conductance $G_{\rm N}$ of the junction and the corresponding resistance 
$R_{\rm N}$ are given by
\begin{equation}
\label{G}
G_{\rm N} = R^{-1}_N = \frac{4e^2}{h} \sum_{n=0}^{\infty} \tau_{n} .
\end{equation}

\emph{Superconducting I-V characteristics.--}
We now turn to the analysis of the current-voltage characteristics in the case in which 
the electrodes are in the superconducting state. As explained above, the current in
the short-junction limit ($L \ll \xi$) can be expressed as a sum of independent channel
contributions as follows
\begin{equation}
\label{iv-dist}
I(V,t) = 2 \sum_{n=0}^{\infty} I(V,t,q_{n}) ,
\end{equation}
where $t$ is the time and $I(V,t,q_{n})$ is the single-channel current of a superconducting
point-contact of transmission $\tau_{n}(q_n)$ (cf.~Eq.~\ref{Tna}). We compute this 
single-channel current using the Hamiltonian approach described in Ref.~\onlinecite{Cuevas1996}.
Let us remind that in the case of a superconducting 
junction at finite voltage, the current oscillates on time as $I(V,t) = \sum_m I_m(V) 
\exp[i m \phi(t)]$, where $\phi(t)$ is the time-dependent superconducting phase difference 
given by the Josephson relation $\partial \phi(t) / \partial t = 2eV/\hbar$. We only consider 
here the dissipative dc current, which we shall simply denote as $I$ from now on. Moreover, 
in the limit $L \ll W$ considered here, one can replace the sum over $n$ in the previous 
equation by an integration. 

\begin{figure}[t]
\begin{center}
\includegraphics[width=\columnwidth,clip]{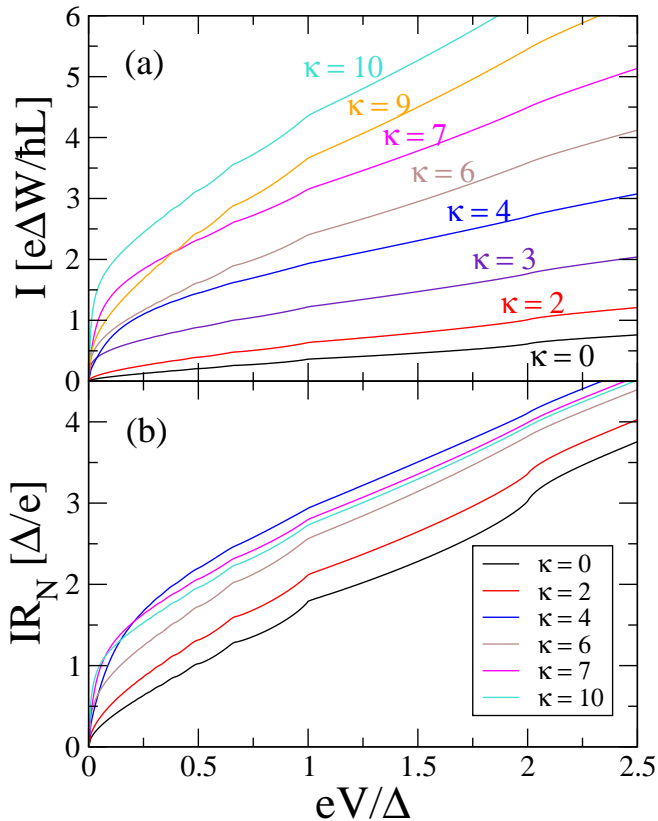}
\caption{\label{iv} (Color online) (a) Zero-temperature current-voltage characteristics
of a superconducting ballistic graphene junction (length L short compared to the width
W and the superconducting coherence length $\xi$), for different values of the gate voltage
in the normal region ($\kappa = e |V_{\rm gate}|L/ \hbar v$). (b) The same as in panel (a),
but the current is normalized by the resistance of the junction in the normal state $R_{\rm N}$.}
\end{center}
\end{figure}

We show in Fig.~\ref{iv} the zero-temperature I-V characteristics for different values of 
the gate voltage. Notice that in the upper panel the current is expressed in an absolute scale,
while in the lower one it is normalized by the normal-state resistance that scales as $R_{\rm N}
\propto L$. At zero gate voltage, i.e. at the Dirac point, the current is formally identical
to that of a short diffusive SNS junction~\cite{Bardas1997}. This is easy to understand with 
the help of Eq.~\ref{Tna}. From this expression one can show that for $V_{\rm gate}=0$ the
transmission distribution adopts the form $\rho(\tau) = (W/\pi L) 1/2 \tau \sqrt{1-\tau}$, which
corresponds to the well-known bimodal distribution for diffusive wires~\cite{Nazarov1994}.
Away from the Dirac point, the non-linearities of the I-V curves are modulated by the gate
voltage. In particular, for gate voltages in which most of the open channels have a high
transmission (see Fig.~\ref{setup}b), the I-V curves are rather smooth (see for instance the curve
for $\kappa=4$). Notice also that contrary to the normal state, the current at certain 
voltages below the gap is not neccesarily a monotonously increasing function of the gate
voltage. This peculiarity is due to the fact that the subgap current is a very non-linear 
function of the transmission coefficients.

\begin{figure}[t]
\begin{center}
\includegraphics[width=\columnwidth,clip]{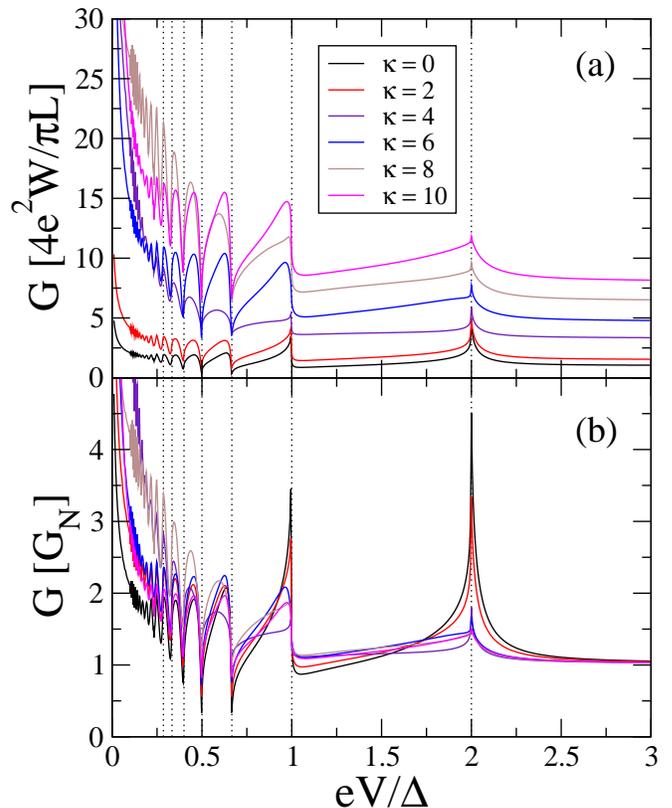}
\caption{\label{cond} (Color online) Differential conductance corresponding to the I-V curves
of Fig.~\ref{iv}. In panel (b) the conductance is normalized by the normal-state conductance
$G_{\rm N}$. As a guide for the eyes, the dotted vertical lines indicate the position 
of the voltages $2\Delta/ne$ ($n=1,2, ...,7$).}
\end{center}
\end{figure}

The non-linearities in the current are better observed in the differential conductance, which
is shown in Fig.~\ref{cond}. The conductance exhibits a very pronounced subgap structure,
although the maxima not always appear exactly at the voltages $2\Delta/ne$ ($n$ integer). As 
explained in the introduction, these peaks are due to the opening of new MAR processes at
those voltages~\cite{Klapwijk1982}. The pronounced peaks (especially at $2\Delta/e$ and $\Delta/e$) 
come from the contribution of the low-transmitting conduction channels. This explains why this 
structure is more pronounced close to the Dirac point (see Fig.~\ref{setup}b). At very low bias, 
the conductance exhibits a square-root singularity ($1/\sqrt{V}$), which originates from the 
contribution of the highly-transmitting channels~\cite{Bardas1997}. In an actual experiment,
this singularity would be masked by the transition to the supercurrent branch at low bias.

\begin{figure}[t]
\begin{center}
\includegraphics[width=0.8\columnwidth,clip]{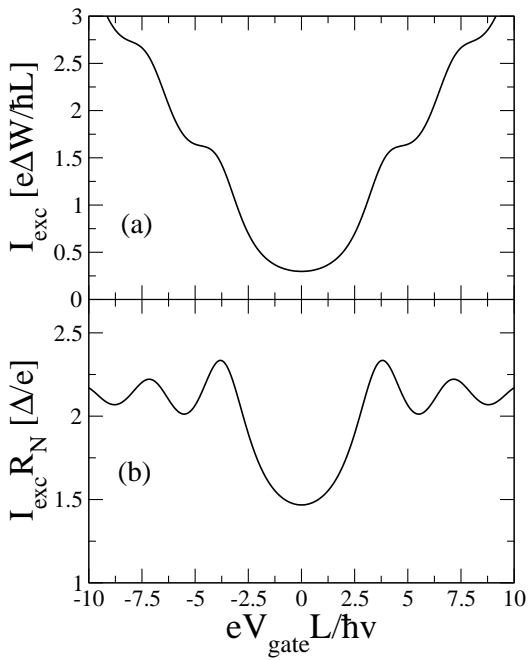}
\caption{\label{excess} (a) Zero-temperature excess current as function of the gate voltage
for a short ballistic graphene SNS junction. (b) The same as in panel (a), but normalized by the
resistance of the junction in the normal state $R_{\rm N}$.}
\end{center}
\end{figure}

Another interesting quantity for a possible comparison with experiments is the so-called
excess current, $I_{exc}$. At voltages $eV \gg 2\Delta$ the current can be expressed as
$I(V) = I_{\rm N}(V) + I_{exc}$, where $I_{\rm N}((V)$ is the current in the normal state 
and $I_{exc}$ is the excess current, which is independent of the voltage. In the case of 
a single-channel point-contact, the zero-temperature excess current as a function 
transmission coefficient $\tau$ can be written as~\cite{Cuevas1996}
\begin{equation}
I_{exc} = \frac{2e\Delta}{h} \frac{\tau^2}{1-\tau} \left[
1 - \frac{\tau^2}{2(2-\tau)\sqrt{1-\tau}} \ln \left(
\frac{1+\sqrt{1-\tau}}{1-\sqrt{1-\tau}} \right) \right] .
\end{equation}
Averaging this expression with the transmission function for the graphene junction, as
indicated in Eq.~\ref{iv-dist}, one obtains the excess current shown in Fig.~\ref{excess}.
The gate modulation of $I_{exc}$ is very similar to the modulation of the critical current
found in Ref.~\onlinecite{Titov2006}. Again, at the Dirac point one recovers the result
of a short diffusive SNS junction~\cite{Artemenko1979,Bardas1997}, i.e. $eI_{exc}R_{\rm N}/
\Delta = (\pi^2/4) - 1$.

\emph{Discussion and conclusions.--}
In this work we have focused our attention in the analysis of the I-V curves of
short SNS graphene junctions, but the idea of combining the transmission function
of Eq.~\ref{Tna} with the single-channel point-contact results can be also used
to study other transport properties of these junctions such as shot 
noise~\cite{Cuevas1999,Naveh1999}, Shapiro steps~\cite{Cuevas2002}, photon-assisted 
tunneling~\cite{Cuevas2002} or even full counting 
statistics~\cite{Cuevas2003,Johansson2003}. For the special case of undoped graphene 
($V_{\rm gate}=0$) the transport properties are identical to those of short diffusive SNS
systems, which have been already reported in the literature~\cite{Naveh1999,Johansson2003,
Cuevas2004}. Moreover, following the recent work of Katsnelson~\cite{Katsnelson2006}, a 
similar idea could be used to study superconducting contacts involving bilayer graphene.

In summary, we have shown that the I-V characteristics of a short ballistic graphene SNS 
junction exhibits a very rich and gatable subharmonic gap structure. This very pronounced 
and sensitive structure predicted here can be straightforwardly compared with 
experimental results in submicron scale junctions~\cite{Heersche2006} and it might constitute 
an ideal test quantity to confirm some of the peculiar consequences of the interplay of
superconductivity and relativistic quantum dynamics.

\emph{Acknowledgments.--}
We acknowledge discussions with Hubert B. Heersche, Pablo Jarillo-Herrero, 
Sebasti\'an Bergeret and Elsa Prada. This work has been financed by the Spanish CYCIT 
(contract FIS2005-06255) and by the Helmholtz Gemeinschaft (contract VH-NG-029).

%&&&&&&&&&&&&&&&&&&&&&&&&&&&&&&&&&&&&&&&&&&&&&&&&&&&&&&&&&&&&&&&&&&&&&&&&&&&&&

\end{document}